\documentclass{kapproc} 
\usepackage{procps} 
\usepackage[dvips]{graphicx}
\upperandlowercase
\kluwerbib 
\begin{document}
\articletitle{The Recurrent Nature of Central Starbursts}
\author{Curtis Struck}
\affil{Dept. of Physics and Astronomy\\
Iowa State University, Ames, IA, USA}
\email{curt@iastate.edu}

\begin{abstract}
New hydrodynamic models with feedback show that feedback driven
turbulence and subsequent relaxation can drive recurrent starbursts,
though most of these bursts fizzle due to premature, asymmetric
ignition. Strong bursts are terminated when the turbulence inflates
the multiphase central disk. The period between bursts is about twice
a free-fall time onto the central disk. Transient spirals and bars are
common through the burst cycle.
\end{abstract}


There has been much discussion at this meeting on triggering
mechanisms for starbursts. This topic is especially interesting when
galaxy interactions, bar-driven inflows and other obvious sources of
triggering are absent. Two other interesting questions are what turns
off starbursts, and are starbursts naturally recurrent given an
adequate gas supply? And finally, if they are recurrent, is the
process deterministic or stochastic?

There are many candidate mechanisms for terminating bursts (see review
of Leitherer, 2001), including 1) gas consumption, 2) gas loss to the wind,
3) conversion of cold gas to warm/hot phases, and 4) inflation of the
central disk without total conversion to hot phases (as in dwarf
galaxy models). The last two of these allow recurrence, the first two
don't. Since there are many complex processes involved, with
incomplete sampling in any one observational waveband, it is hard to
assemble a complete picture. Many models have been made of particular
parts of the coupled starburst + wind phenomena.

Although models for all the thermo-hydrodynamic aspects of the
starburst/wind phenomena are not possible, exploratory models of some
the important couplings can shed light on the questions above. As a
step towards that goal, I present here some preliminary results of
numerical hydrodynamical models. These are N-body/SPH models carried
out with the Hydra3.0 code of Couchman, Thomas, and Pearce (1995),
which includes optically thin radiative cooling (Sutherland and
Dopita) for temperatures above $10^4 K$. The model galaxies consist of
three components: a dark matter halo (10,000 collisionless particles),
stellar disk (9550 collisionless particles), and a gas disk (9550 SPH
particles). The mass per particle is $6 \times 10^6\ M_{\odot}$, and
the total galaxy mass is $1.75 \times 10^{11}\ M_{\odot}$.

A feedback prescription was also added to the Hydra code. A gas
particle is marked as star-forming in this prescription when its
temperature is less than a threshold value of about $10^4\ K$, its
density is above a threshold value of about $0.14\ cm^{-3}$, and these
thresholds are not re-crossed during a time of about $10^7\ yrs$. The
latter condition is a computationally cheap way of insuring that the
cold, dense region is likely bound. The small value of the density
threshold is a symptom of the modest particle resolution; regions that
are much denser are rarely resolved, but we assume they exist within
the densest regions.

Once the heating is initiated it is maintained for $10^7\ yrs$, and
the particle internal energy is increased by 10\% in each timestep
(typically about $10^5\ yr$) until a maximum temperature of about
$10^6\ K$ is reached. This corresponds to an energy input of $1.1
\times 10^{54}\ ergs$ per star-forming particle. (This feedback
formulation is discussed in more detail, and compared to others Smith,
2001.)

If we assume that a typical supernova injects about $10^{51}\ ergs$
into the gas, then about 1100 SN are needed to generate this feedback
energy. If we assume that the mass of a typical SN is about $10\
M_{\odot}$, and adopt a Salpeter mass function over a range of $0.2 -
100\ M_{\odot}$, then about 10\% of the stars are SN progenitors, and
we need to form a star cluster of mass $\simeq 10^5\ M_{\odot}$ to
obtain the needed energy. This is about 2\% of the gas particle
mass. Gas consumption is not included in these models, or equivalently
we assume instant replenishment of the (small) losses.

These energy estimates do not include losses due to cooling (until the
heating is terminated).  However, the feedback heating should not be
directly equated to thermal energy. It is also an algorithm for
inducing increased kinetic energy and mass motions via local pressure
effects, which begin at a "sub-grid" scale. Once the heating is turned
off, the affected particles can cool rapidly due to both adiabatic
expansion and radiative cooling. The net pressure generated by the
feedback, and its affects on surrounding particles seem generally
realistic, even if the thermal details are unresolved. In the future,
these details must be resolved to determine quantities like the mass
fractions and scale heights of different thermal phases.

The first result is that with sufficient gas supply, and reasonable
feedback parameter values, the models are intrinsically bursty at a
moderate level, see Figure 1. The initial model was allowed to relax
with an adiabatic equation of state plus cooling, but without
feedback. The feedback was turned on at the start of these runs, which
resulted in some large amplitude bursts.  After that the models
settled down to a more "steady, bursty" character with the following
characteristic properties. 1) The largest bursts are roughly periodic,
with a period of a bit less than 100 Myr. This period is much longer
than the time delays of the feedback model. 2) The amplitude of the
largest bursts varies substantially. The horizontal line in the figure
serves as a useful guide. The case shown by the dashed curve has
about an equal number of burst peaks above this line and just below
it. The case shown by the solid curve has bursts that peak
significantly below the line for most of a Gyr, then several bursts
that peak above it, including the double burst between the vertical
lines. 3) The burst durations are typically about 10 Myr (like the
feedback time delay), but double or multiple echo bursts are not
uncommon.

\begin{figure}[ht]
\scalebox{0.4}{\includegraphics{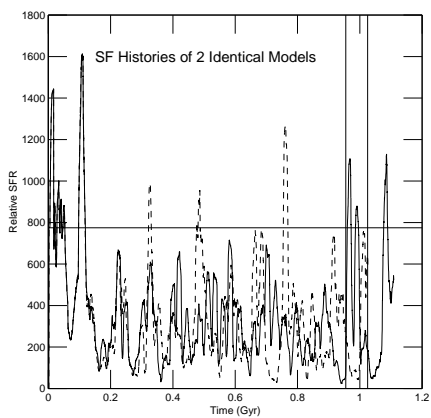}}
\caption{The number of actively heating (feedback) particles, assumed
proportional to the SFR, as a function of time in two models with
essentially identical initial conditions. The horizontal line in the
figure is drawn at the level of the mean + 2 standard deviations,
see text.}
\end{figure}

I have analyzed the bursts that peak above the $2\sigma$ horizontal
line, but excluding the initial transients, in more detail. Adopting
the star formation (SF) mass per particle and IMF described above, the
stellar mass produced in these bursts ranges from $4-50 \times {10^6}\
M_{\odot}$, with a mean of $30 \times {10^6}\ M_{\odot}$. The
corresponding SFR ranges from $1.6-3.6\ M_{\odot}/yr$.  These seem to
be fairly reasonable values for an isolated core starburst in a
late-type, and so support the parameter values used in the feedback
algorithm. However, a number of questions remain.

What determines the burst period? The burst period is slightly less
than twice the free-fall time of a particle at 2-3 kpc above the disk
plane, or the time for a boosted gas particle to travel to the top of
its trajectory and return to the plane, as observed in the models.

Why do the burst peaks vary so much? Viewed from above, the nuclear
disk gas is usually found to be concentrated in flocculent spirals,
and transient, rather irregular bars. After a starburst large bubbles
and voids often appear, and the spirals and bars are disrupted or
rearranged. A great deal of turbulence is generated. The next burst is
more likely to be of large amplitude if these waves reform
symmetrically and transfer a relatively large amount of gas to the
center. More frequently, these waves develop asymmetrically, and an
off-center gas concentration triggers a SF hot spot prematurely. This
often leads to some propagating SF, but destroys the chance for a
large burst. In the case, shown by the solid curve in Fig. 1,
successive ``fizzles'' extend the time between large bursts to nearly
1.0 Gyr.

Where's the wind? Observations indicate that the wind mass is roughly
equal to the mass of stars produced in the burst (see Strickland,
2004). The hot wind is not resolved in these models. The models do
suggest that the SF that generates it occurs in patchy
concentrations. Thus, the wind is probably a sum of local gusts. This
seems in accord with recent observations of M82 and Arp 284 presented
at this meeting and observations of other starburst galaxies.

What about all the gas boosted out of the plane? First of all, there
is observational evidence indicating that substantial masses of gas
are kicked out over kpc distances by bursts. This evidence includes
large HI scale heights in M82 and NGC 2403 (Fraternali et al. 2004),
extended dust distributions in edge-on galaxies like NGC 891 (Howk \&
Savage 2000), evidence that molecular clouds are broken down but not
destroyed in starburst regions (Gao \& Solomon, 2004), and the small
filling factor of the hot gas in wind galaxies (Strickland, 2004).
The high latitude gas in the models has a wide range of thermal
phases.

In sum, the simulations suggest a mosaic model for core starbursts,
with the following properties.  1) Star clusters form in the densest
regions, create hot spots, which may eventually break out as wind
gusts. 2) Hot spots also drive turbulence over a wider area, and can
propagate the SF. 3) Eventually, central regions become so turbulently
stirred, shredded and puffed up that SF crashes.  4) If not too much
gas is consumed or blow out, clouds reform and generate recurrent
bursts. The models suggest that starbursts are naturally recurrent. We
have not yet undertaken models of bursts driven by rapid mass transfer
or merging, but it seems likely that they can overcome the
fizzle-effect and drive large burst amplitudes.


{}
%
%
\chapappendix{Supporting Figures and Text (for CD-ROM)}

In this appendix I describe two figures that illustrate the
effects of a starburst on the central disk hydrodynamics. The first
figure (A.1) shows face-on views of the central disk (20 x 20 kpc
frames) at 6 times within the range indicated by vertical lines in
Figure 1. The model used was the one corresponding to the solid curve
in Figure 1, and which has a double burst during this period. 

Figure A.1 also shows gas particles that are currently experiencing
heating from young stars as described above. These star-forming
particles are marked by green plus signs. The two SF burst peaks occur
near the times shown in the second and fourth frames of Figure A.1. The
figure shows that the first burst is very concentrated in the
disk center, while the second burst is concentrated in a spiral on one
side of the center. Flocculent spirals are common, as are large holes
in the gas distribution away from the center. At other times, or in
other models, transient bar-like structures can also be seen.

\begin{figure}[ht]
\scalebox{0.8}{\includegraphics{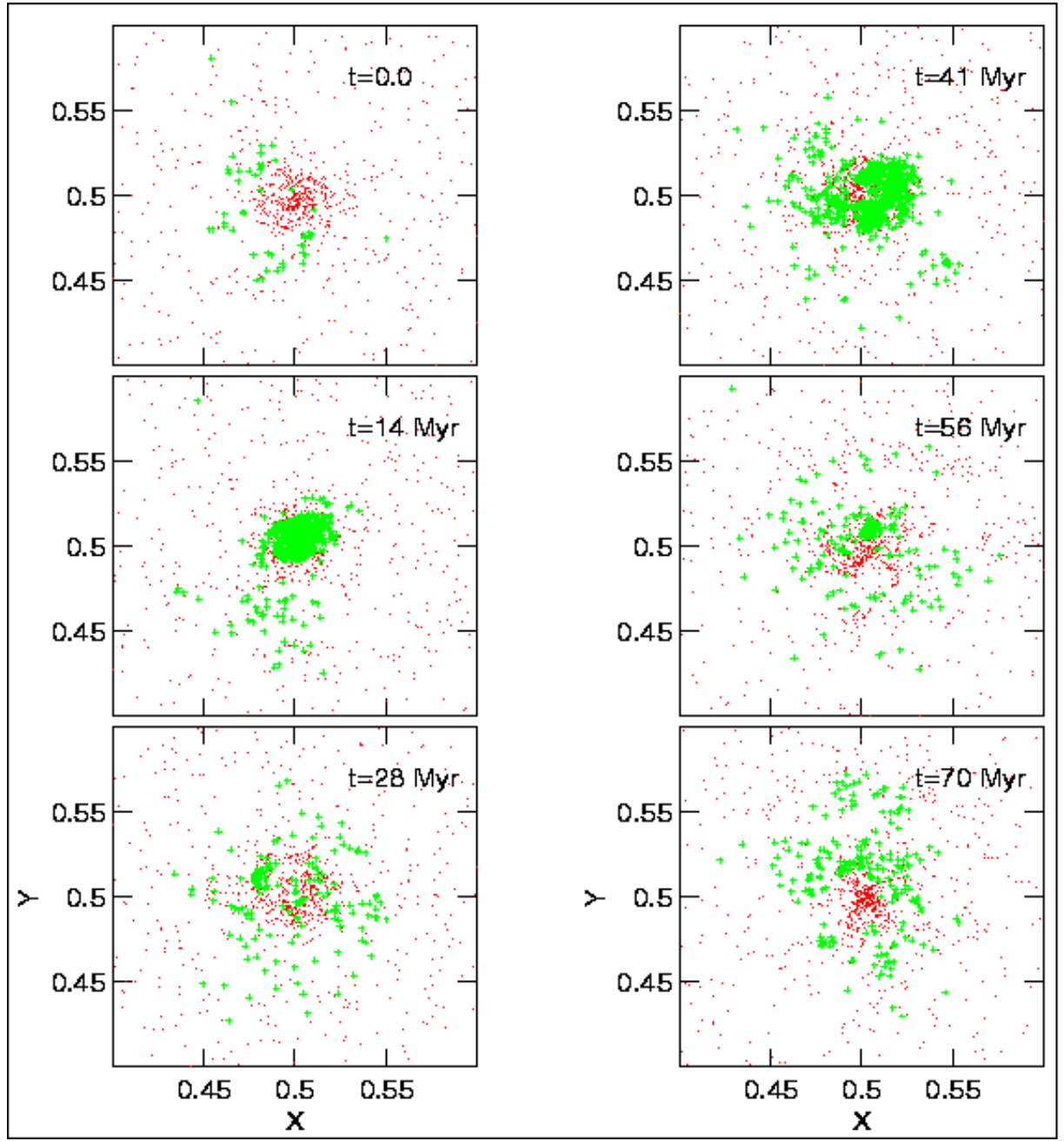}}
\caption{Six snapshots showing the distribution of gas and star
formation during the time of a double starburst, which is bracketed by
vertical lines in Figure 1. The central disk is shown face-on (xy
view) in each image. Red dots are gas particles. Green plusses
mark star-forming gas particles.}
\end{figure}

\begin{figure}[ht]
\scalebox{1.0}{\includegraphics{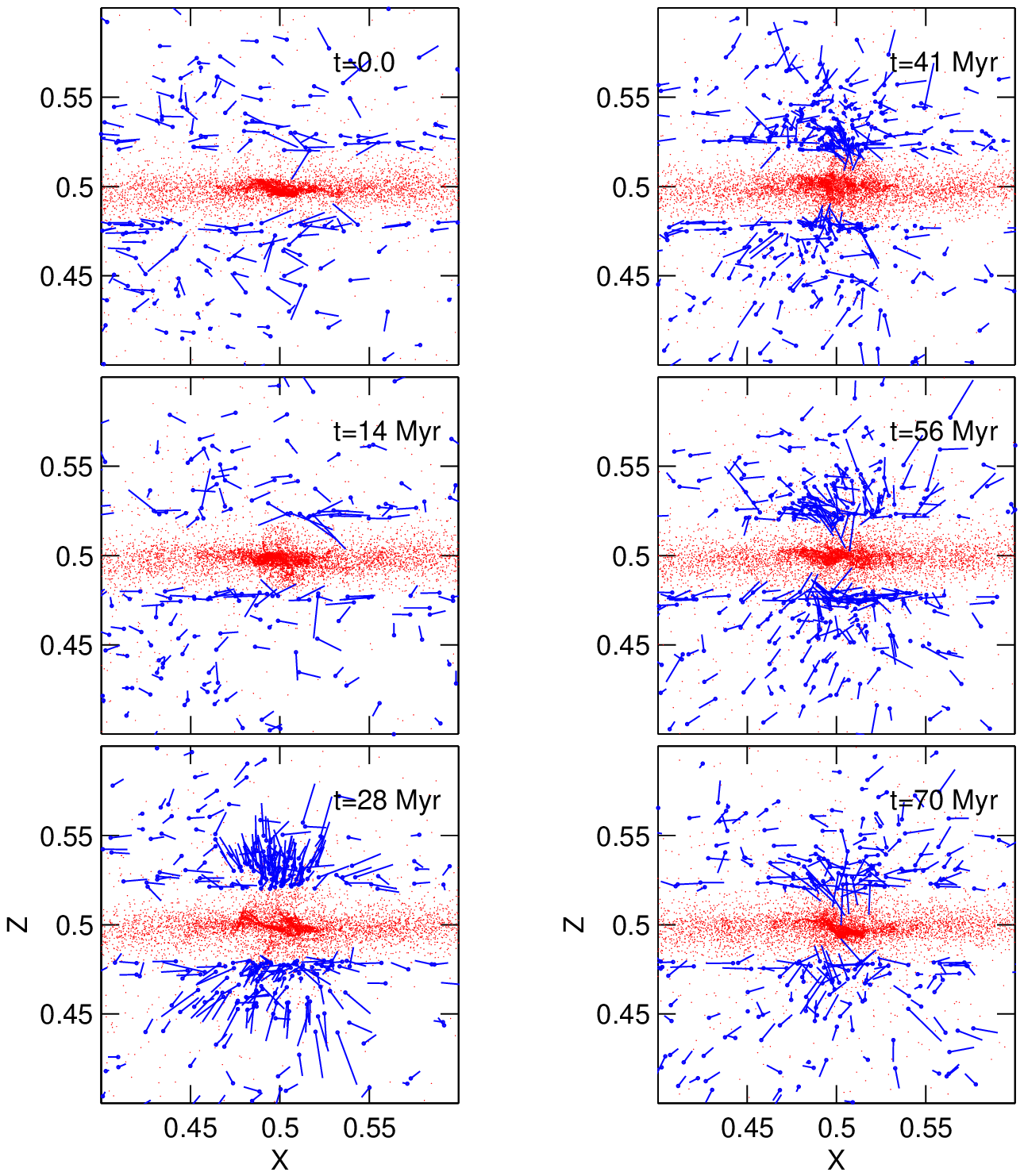}}
\caption{Six snapshots of the gas distribution in views perpendicular
to the disk plane (xz views). The times shown are the same as in the
previous figure. Red dots are gas particles. The x-z velocity
components are shown by blue line segments for every third gas
particle at a distance of more than 0.02 units (2 kpc) from the disk
midplane. The magnitude of the velocity in km/s equals the segment
length in code units times 100. The third panel shows the induced
outflow most clearly.}
\end{figure}

Figure A.2 shows views orthogonal to the disk plane (x-z) at the same
six times, and on the same length scale. The inflation of the central
gas disk by the starbursts can be seen in this sequence. I also show
velocity vectors of selected particles as described in the figure
caption. 

Before the onset of the double burst the central disk had experienced
a period of unusually low SF. This is illustrated by the relatively
small number of gas particles at any significant distance from the
disk plane in the first two frames of Figure A.2.

By the time of the third frame the initial core burst has ended, but
large numbers of gas particles are moving away from the plane with
high velocities. Despite the second burst the coherent outward motion
seems to stall by the time of the fourth and fifth frames, and infall
dominates the sixth frame. 

The ineffectiveness of the second burst my be partly due to the fact
that it is off-center, and can dump some of its energy to nearby holes
or voids within the disk. Off-center or hot spot bursts are more
common than central bursts in these models, and deserve much more
study. 

\end{document}